\documentclass[acmtocl]{acmtrans2m}

\usepackage{epsfig,graphics}
\usepackage{amsmath}
\newdef{definition}[theorem]{Definition}
\newcommand{\two}{\mathit{two}}

\begin{document}

\title{About the efficient reduction of lambda terms}

\author{Andrea Asperti\\
        DISI: Dipartimento di Informatica - Scienza e Ingegneria\\
        Mura Anteo Zamboni 7\\
        40127, Bologna, Italy}

\begin{abstract}
  There is still a lot of confusion about ``optimal'' sharing in the lambda
  calculus, and its actual efficiency. In this article, we shall try to
  clarify some of these issues. 
\end{abstract}

\category{}{Theory of computation}{Lambda calculus}
\category{}{Theory of computation}{Abstract machines}
\category{}{Theory of computation}{Equational logic and rewriting}
\category{}{Software and its engineering}{Functional languages}

\maketitle

\section{Introduction}
In relation to rewriting techniques, {\em sharing} is the ability to avoid
duplication of reduction work, due to duplication of subterms. The issue is 
relatively trivial at first order, but it becomes much more entangled
as soon as we pass to a higher order framework, for which the lambda calculus
provides a paradigmatic example.

Consider the well known beta rule
\[\lambda x.M\,N \to M[N/x]\]
If the argument $N$ gets duplicated and it contains
a reducible expression, its reduction will be duplicated too.

It may seem that an eager strategy (possibly delayed ``on demand'', as
in the ``call by need'' strategy) could solve the job. Unfortunately,
this is not the case.

Let us consider first the case of weak frameworks. In this case, functions
are treated as values and reduction is never pursued under a
$\lambda$-abstraction.
So, if the argument $N$ is a lambda expression containing a redex $R$, and
$N$ is duplicated, the reduction of $R$ will be repeated in each instance.
A typical situation is when the argument $N$ is obtained as a
{\em partial} instantiation of some functional $F$. To make things
very simple, let us suppose $F = two = \lambda xy.x(x\,y)$
(the Church integer) and let us instantiate it with the identity
$I = \lambda x.x$
\[N = two\;I \to \lambda y.I(I\,y)\]
that is a weak normal form. If $N$ gets duplicated, the two internal
applications of the identity will be duplicated too.

This may have {\em very nasty} effects. Consider the following weak reduction
\[
\begin{array}{rl}two\;two\,I & \to two\, (two\,I) \\
  & \to two\, (\lambda y.I(I\,y)) \\
  & \to \lambda y.(\lambda y_1.I(I\,y_1))(\lambda y_2.I(I\,y_2)\,y)
\end{array}
\]
where we renamed variables for the sake of readability.
We have just doubled the number of internal applications of the
identity! If we start with $n$ applications of $two$
\[\underbrace{\two \dots \two}_{n\, \mathit{times}} I\]
we end up with a term containing $2^n$ applications of the
identity and all of them will need to be reduced when the term will be
feed with an extra argument (e.g. an additional identity).

We warmly invite the readers to write and evaluate the term
\begin{equation}
  \label{term1}
  n\,two\, I\, I
  \end{equation}
(where $n$ and $two$ are Church integers) in their favorite (weak) functional
programming language, and observe the exponential explosion of the
complexity when $n$ grows (no matter if the language is lazy or strict,
or if it adopts combinators or closures). On the other side, innermost reduction
of the previous term is just linear in $n$.

So, is rightmost innermost reduction the correct solution?
Of course, not. As a trivial example, consider the term
\begin{equation}
  \label{term2}
  I\,(n\;two)\,I\,I
  \end{equation}
Rightmost innermost reduction would start normalizing $(n\;two)$ that
is the Church integer for $2^n$ and has exponential dimension, hence the
whole reduction would be exponential too.

What happens in (the innermost reduction of) example $(\ref{term2})$ is that the term $I$ inside $\lambda y.I\,(I\,y)$
of example $(\ref{term1})$ is replaced by
a local variable, postponing the instantiation with the identity
to a later stage. That is to say, that is not the duplication of
redexes that matters, but the {\em unnecessary, blind duplication of
  applications}. For instance, with environment machine, any time
we open a closure and the internal code contains an application, we
are possibly duplicating reduction work.

\begin{figure}[htb]
  \begin{center}
  \includegraphics[width=.7\columnwidth]{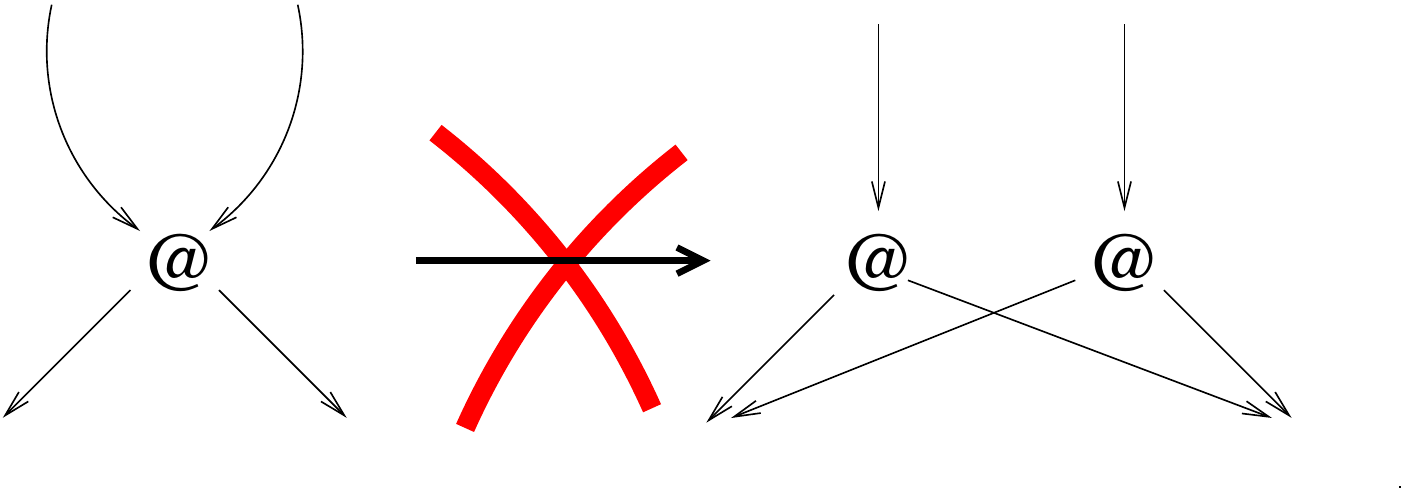}
  \caption{Forbidden duplication of applications}
  \label{fig:appl}
  \end{center}
\end{figure}

But applications and lambda abstractions are just dual operators,
so is the duplication of lambda abstractions dangerous too, from the
point of view of sharing?

\begin{figure}[htb]
  \begin{center}
  \includegraphics[width=.7\columnwidth]{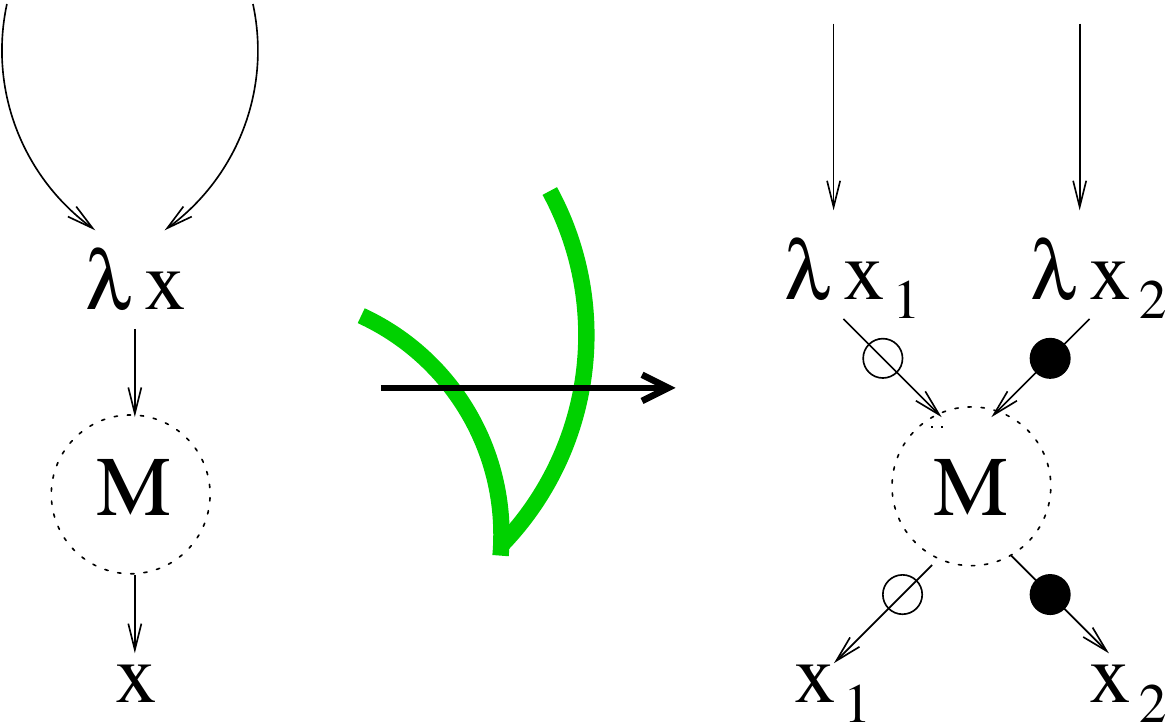}
  \caption{Legal duplication of $\lambda$-abstractions}
  \label{fig:lam}
  \end{center}
\end{figure}

In principle, no, it is not. The point is that if the abstraction
node is shared, there are already two (or more) {\em different}
calls to the function, that will give rise to {\em different} redexes.
The big challenge, however, is to duplicate the abstraction node
{\em without jointly duplicating the whole body} of the function
(that could contain applications). The really delicate part is to understand
what happens at the level of variables, since they can now be
bound by one or the other of the two abstractions, requiring some
form of ``unsharing'' (see Figure \ref{fig:lam}). 
The correct management of sharing and unsharing is not trivial.
It was solved for the first time by Lamping \cite{Lamping90}, and later
revised and improved by many other people. One usually refer to
this part of the algorithm as ``bookkeeping'' work, to distinguish
it from duplication work and the actual firing of $\beta $-redexes.

Let us also observe that, in the terminology of interaction nets \cite{Lafont90},
the different behavior
between the duplication of applications and lambda abstractions
resides in the fact that in the latter case (Figure~\ref{fig:lam})
duplication is requested
at the principal port of the node, while in the case of the application
(Figure~\ref{fig:appl}), it is requested at an auxiliary port.

\section{Reduction by families}
L\'evy developed the theory of optimality long before an implementation
for it was available (in fact, the problem remained open for quite
a long time). The precise definition of optimal sharing is not simple,
and we shall postpone it for a moment.
Two redexes that are sharable according to L\'evy are said to
belong to a same family, and optimal reduction is simulated on lambda
terms by firing ``in parallel'' all redexes in a same family. Family
reduction has very nice properties: the most interesting one is that
it satisfies a one-step diamond property. As a consequence, as far as we reduce
needed redexes, the length of a normalizing reduction (if it exists)
does not depend on the strategy. This fact supported the conjecture that
family reduction could provide an interesting measure of the
``intrinsic complexity'' of lambda terms, i.e. the cost required to
compute the normal form of a lambda term independently from the
reduction technique.

Before addressing this issue, let us consider a different, simple 
reduction technique: parallel $\beta$-reduction in Takahashi's sense
\cite{parallel}, that allows us to fire in parallel (in a single
step) all redexes in a given term. Clearly, this is a superoptimal
reduction technique: all redexes in a L\'evy's family are parallel
in Takahashi's sense, but non all parallel redexes eventually belong
to a same family (that is, not all of them are sharable).

The potential parallelism inherent in $\lambda$-terms can be
very easily understood by restricting the attention to the simply typed
case (the following argument was spelled out for the first time in the
appendix to \cite{AspertiL13}).

Working with simple types, it is traditional to define a notion of
{\em degree} of a redex $R$ in the following way (see e.g.\cite{GirardJY:prot}).

\begin{definition}[degree]
The {\em degree} $\partial(T)$ of a type $T$ is defined by:
\begin{itemize}
\item $\partial(A) = 1$ if $A$ is atomic
\item $\partial(U \to V)= \mbox{max}\{\partial(U),\partial(V)\} + 1$
\end{itemize}
The {\em degree} of a redex $(\lambda x:U.M)N$ is $\partial(U \to V)$,
where $V$ is the type of $M$.\\
The {\em degree} $\partial(M)$ of a term $M$ is the maximum among the degrees of
all its redexes.
\end{definition}

A crucial property of the simply typed
lambda calculus is that a redex $R$ of type $U \to V$ may only create redexes
of type $U$ or of type $V$, hence with a degree strictly less than that
of $R$. As a consequence, each simply typed lambda term $M$ can be reduced to
its normal form with a number of parallel reduction steps bound by its
degree $\partial(M)$. On the other side, we can encode complex 
(arbitrarily large Kalmar-elementary) computations in $\lambda$-terms
with low-degrees
(see \cite{May74,St77}). So, this two facts together prove
that the {\em amount of a parallelism} in $\lambda$-terms
is not elementary recursive.

Does this say anything bad about parallelism? No. On the contrary,
there is a {\em huge} amount of parallelism in lambda terms
(more than one could have expected), so it seems to be rather a good idea
to try to exploit it. Of course, the speed up we may expect is never
larger then the degree of parallelism, and if it is finite (or
even elementary in the size of the term!) the execution of
large elementary computations (with an exponential height larger
than that of the available parallelism) will remain elementary. 

Coming back to optimality, the important result proved in 
\cite{AspertiMairson} was that 
most of these parallel redexes are actually {\em sharable} in L\'evy's
sense, so that, again, you may reduce a simply typed lambda term
in a number of family reductions that is approximately linear in
its size (!!). Technically, this implies that (on a sequential machine)
the cost of {\em sharing} a single redex cannot be bound by any
elementary function, but this is merely due to the {\em enormous amount
  of sharing} that is inherent in lambda terms.

Stated in another way, we already concluded that parallel reduction
does not look a bad idea. Then we discovered that most of the parallel
redexes can be actually shared, that looks like an even better idea: why
wasting parallelism by duplicating work if you can share it? However,
the amount of sharing can be so - inconceivably - large
that (in worse, pathological cases) cannot
be handled in elementary time in the size of the term. That's all.

The result in \cite{AspertiMairson} tells you {\em nothing} about the
efficiency of optimal reduction. The surprising result is that in lambda terms,
due to higher order, we have {\em much more} sharing (in L\'evy's sense) than
expectable.
As a consequence:
\begin{itemize}
  \item the computational cost per family may be huge
  \item the length of family reduction is not a good measure of the intrinsic
    complexity of terms
\end{itemize}
    
\section{Efficiency, in theory}
Intuitively, sharing graph reduction \`a la Lamping performs the {\em minimum
  amount of duplication} required by the computation. However, as we already
explained, in addition
to this duplication work, there is also an additional ``bookkeeping'' work
required to enforce the correct matching between sharing and unsharing.
This is usually implemented by means of different {\em levels} of sharing,
and the introduction of suitable operators acting as
{\em brackets} in the graph to delimit the scope of duplicators,
dynamically changing their levels. This part of the algorithm is pretty
complex, and its cost is not so clear yet. In particular, as proved in \cite{AspertiC97}, if you are not careful in the management of brackets, they can easily
accumulate, resulting in an exponential overhead. For instance, Gonthier's
implementations \cite{GonthierAL92popl,GonthierAL92lics} are just {\em wrong}, from this respect.

The accumulation problem described in \cite{AspertiC97},
was not present in Lamping's original algorithm \cite{Lamping90},
neither in the Bologna Optimal Higher Order Machine (BOHM) \cite{AspertiGN96},
or in later implementations such as Lambdascope \cite{lambdascope}.
It is conjectured that bookkeeping only adds a polynomial overhead to
the reduction cost, but there is no proof of this fact.

To avoid to take bookkeeping into consideration, it was natural to look
for frameworks where there is no need for it. A particularly interesting
case was provided by {\em elementary linear Logic} \cite{Girard98}, that is
a logic with boxes but no dereliction, expressive enough to code all
elementary functions. The sharing graph reduction of lambda terms
typable in elementary linear logic can be done without the use of
brackets, and hence without bookkeeping.

Rephrasing \cite{AspertiMairson} in this context, \cite{AspertiCM04} showed that
the non elementary cost of optimal reduction is not due to bookkeeping
(which one may suspect to add superfluous work), but to the
(apparently unavoidable) duplication work. If you accept
the fact that optimal reduction performs the minimal amount of duplication,
you will have {\em at least} the same operations, and hence the same
computational cost in {\em any} reduction technique.

The efficient nature of optimal reduction in absence of bookkeeping
was confirmed by \cite{BaillotCL11}, who considered a class of $\lambda$-terms
of known bounded complexity
(polynomial and elementary time) and investigated the cost of their normalization via
sharing graphs: the cost stays in the expected complexity class.

More recently, still working in a ``bookkeeping free'' framework, and making
a direct syntactical comparison with a standard graph rewriting machine,
\cite{SolieriDICE} showed that sharing graphs can only improve performances.

In conclusion, while there are several examples of classes of lambda terms where optimal
reduction outperforms standard techniques, there is so far no known
counterexample to its computational efficiency. 

\section{Efficiency, in practice}
So, if optimal reduction is so good, and apart from the benighted
ostracism of traditional schools, why functional programming languages
are not yet implemented in this way?

First of all, we should make a distinction according to the intended
use of the normalization algorithm. There are essentially two different
settings where
normalization of $\lambda$-terms plays a role: the first one is in
higher order logical frameworks based on Martin-L\"of type theory
(e.g. for type-checking of
dependent types, or when deploying {\em reflection}); the second setting
is as core of real functional languages. We shall discuss them
separately.

\subsection{Higher order logical frameworks}
\label{sec:hol}
The most important use of reduction in this context is to check
convertibility of $\lambda$-terms: since the calculus is confluent
and normalizing, two terms are convertible if and only if their
normal forms are equal. However, this is just an {\em extrema ratio}:
there is no evidence at all that the best way to check convertibility
is via normalization, and in fact, up to our knowledge, no logical
framework implements it in such a brute force way.
In the vast majority of cases,
two terms are convertible just because are equal (even if not normal),
and it would be a major waste of time to normalize them. Even if
they are not equal, they could just be few reduction steps afar (e.g.
one could be obtained by the other by folding/unfolding a few definitions).
In this case, the use of suitable convertibility heuristics, or a tighter
control of constant unfolding could be substantially more beneficial
than improving the efficiency of reduction.

In the case of optimality, the use of normalization for comparing
terms poses a few additional problems, since there is the need to
{\em inspect} the normal form\footnote{Note that no functional programming language gives you the ability to inspect higher order values, e.g. you cannot read back a closure: this is just an issue
  for convertibility.}. This can be done in two ways:
either by traveling in the resulting graph, computing {\em paths}
in it, or via a {\em readback} procedure that reconstructs the
$\lambda$-term out of the graph. At present, no precise
bound at the complexity of these operations is known, but they do not look
too complex. The delicate point is that, in this case, it does not make sense
to compute complexity in terms of the size of the input, since
a small sharing graph may result in a huge lambda term \cite{LawallM96}. It is
conjectured that, starting from a sharing graph {\em in normal form},
the complexity of the readback procedure is just linear in the size
of the {\em resulting} term (that, for the sake of comparing term, is
the best we may expect), but there is no proof of this fact.

Reduction is also a key ingredient of the reflection
technique \cite{Ring,BarendregtB02},
whose basic idea is to check a property by running a suitable {\em certified}
decision procedure. For instance, in order to compare two regular
expressions, we can build the corresponding automata and execute a 
bisimulation algorithm over them. In this case, having an efficient way of
evaluating lambda expressions may be important; however,
for the most typical uses of reflection, and especially
for small scale reflection \cite{ssreflect}, optimal reduction looks
a bit overkilling.

There is a final point that, at present, may advise against the adoption
of optimal reduction in logical frameworks. Reduction is one of the
most primitive operations in higher order logical frameworks, and
a basic component of the type-checking/verification algorithm.
So, it is part of the so called {\em kernel} of these systems: a
component whose correctness must be trusted. To this aim, it
has been argued that kernels should be small (in terms of lines of code),
in order to improve confidence in their
implementation\footnote{This conception is possibly a bit outdated. Instead
  of having a {\em small} kernel, it would be better to have a {\em verified}
  kernel, of course, no matter what its size could be.}. While it is possible to implement abstract reduction
machines for lambda terms in a few lines, sharing graphs eventually require
a bit more code, and maybe it is not such a good idea to try to put
this machinery in the kernel.

\subsection{Functional programming}
The first issue to face, when considering optimal reduction for the implementation
of a real functional programming language, is to understand if the
technique can be generalized to a larger and more flexible
calculus (coding everything as pure lambda terms is, of course,
not a feasible solution). Since sharing graphs can be expressed in
terms of interaction nets, the natural idea is to generalize the
logical operators from the application-lambda abstraction pair, to
a generic setting of (higher order) interaction operators. This
naturally leads to {\em interaction system} \cite{AspertiL94},
that are the elegant synthesis between interaction
nets and Klop's higher order combinatory reduction systems \cite{Klop80}.
Interaction nets are expressive enough to cover all inductive data
structures, primitive fix-points and recursion, and also effective
numerical computations where each integer is treated as a different constructor
processed via primitive arithmetical operations.
Interaction system can be implemented by means of sharing graphs with
no additional burden with respect to lambda-calculus \cite{AspertiL96},
demonstrating that sharing graphs just provide the abstract machinery
for dealing with (optimal) sharing in a higher-order setting, independently
from the rewriting rules.

\begin{figure}[htb]
  \begin{large}
\begin{center}
\begin{tabular}{|cc|}
  \hline
  {\bf first order} & {\bf higher order}\\
  \;\;direct acyclic graphs (dags)\;\;& \;\;sharing graphs\;\;\\
  \hline
\end{tabular}
\caption{Sharing machinery}
\end{center}
\end{large}
\end{figure}\vspace{-.3cm}
  
The Bologna Optimal Higher-order Machines (BOHM) \cite{AspertiGN96}
provided a prototype
implementation of the above ideas. BOHM was written in C, and aimed
to efficiency, in order to compare with real implementations.
Several benchmarks are given in \cite{OptimalBook}. On pure lambda
terms (see pag.296-230) BOHM outperformed both Caml Light and Haskell,
while remaining competitive on typical symbolic computations. On more 
numerical computations Caml Light was sensibly faster (up to one
order of magnitude), that was not surprising due to the underlying
overhead of graph rewriting. 

The main problem we faced when implementing sharing graphs was
not related to performance but to memory consumption. This may look
surprising since the
point of optimality is precisely to be as parsimonious as possible
in the duplication of data structures. However, the two things
have very little in common. In general, there is a well known tension
between time and space: you may improve time by sacrificing space,
and conversely you may save space by spending more time. For instance,
Savitch algorithm for graph reachability (implying PSPACE = NPSPACE)
works in space $O(log^2(n))$ where $n$ is the number of nodes of the graph, but
its time complexity is $O(nn^{log(n)})$; this is to be compared with the best
algorithms in time, that have time complexity $O(n^2)$ (linear in the
size of the graph) but require $O(n\,log(n))$ space.
In many interesting
cases, a data type can be more compactly encoded in terms of a
{\em procedure} producing it\footnote{This is the case for all non
random numbers according to Kolmogorov complexity.}: a zipped file saves space at the cost
of unzipping the information when required.
As another example, all program transformations meant to
improve performance such as inlining, unfolding or loop unrolling
typically augment the dimension of the code.

To make an example relative to sharing graphs, consider
a fixpoint definition
\[F = \Theta\,M \to M\, (\Theta\, M)\]
where $\Theta$ is some fixpoint operator.
An invocation of $F$ will result in a lazy unfolding and partial
evaluation of its body, as required by the computation.
To avoid to repeat work, this unfolded form must be saved as
a new, optimized version of $F$:
\[F = M (M \dots (M\; (\Theta M)))\]
For instance, after invoking a recursive definition of a factorial
function on the number $20$, the new definition of the factorial will look
like a sort of case switch for the first $20$ integers, followed
by a recursive call to deal with the remaining cases. This may look
as a desirable effect (a sort of naive form of memoization), but
in many situations things are not so clear, possibly leading
to a large consumption of memory space. Of course, you may
renounce to share global definitions with their invocation instances,
making local copies instead, but this clearly goes against the very
idea of optimality.

Twenty years ago, this looked like a serious problem; since then,
memory has become
much cheaper and maybe, in the Big Data era we are entering,
this is not a
real issue any more.

\section{super optimal strategies}
\begin{figure}[htb]
  \begin{center}
  \includegraphics[width=.7\columnwidth]{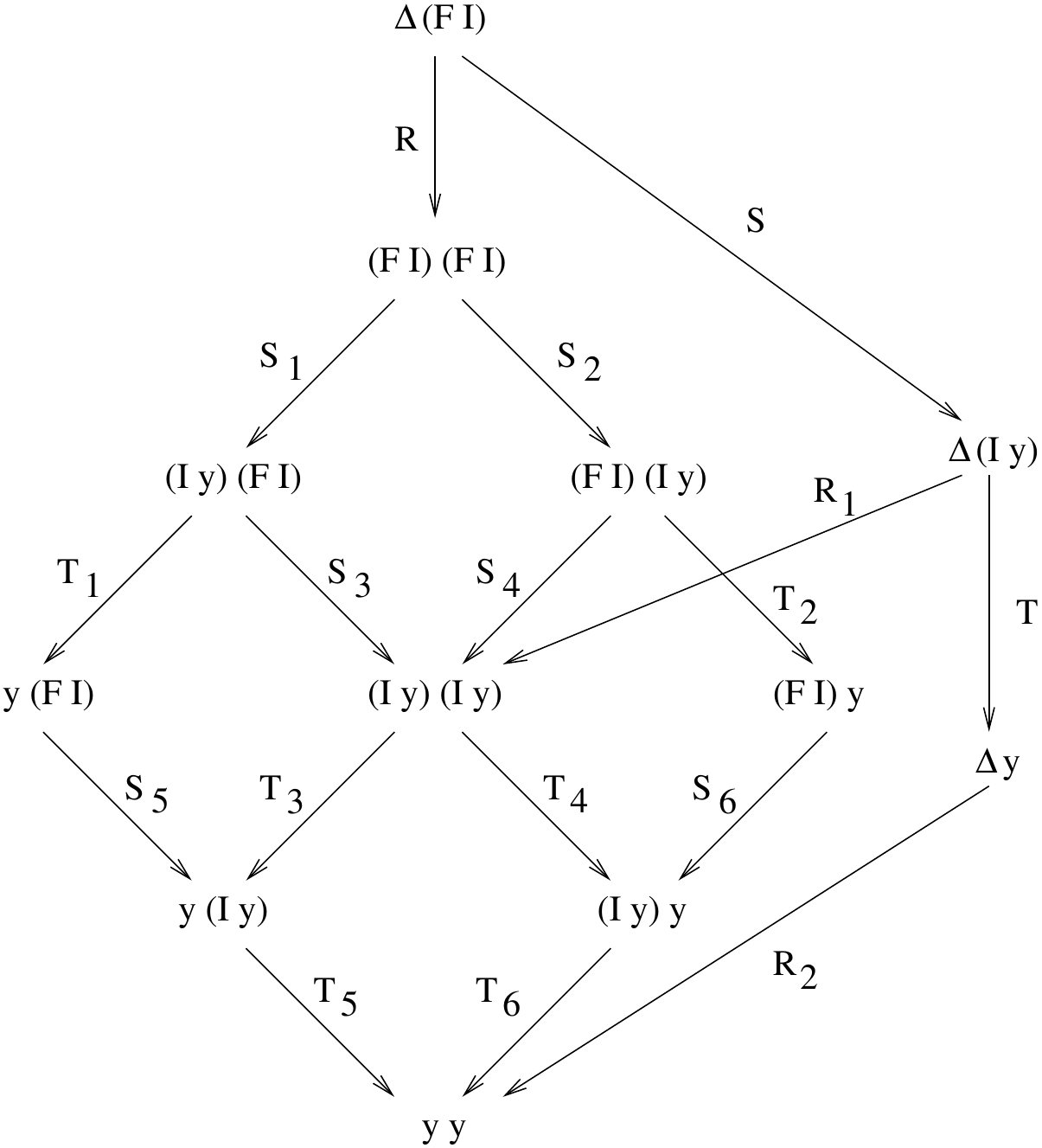}
  \caption{$\Delta = \lambda x.x\,x$, $F=\lambda z.z\,y$ and $I=\lambda x.x$}
  \label{fig:esempio}
  \end{center}
\end{figure}

To address the possibility to have super optimal reduction techniques
for lambda terms we need to better understand the definition of
optimal sharing according to L\'evy.
Let us start with an example. Consider the development for the term $M=\Delta(F\,I)$
described in Figure~\ref{fig:esempio}, where
$\Delta = \lambda x.x\,x$, $F=\lambda z.z\,y$ and $I=\lambda x.x$.
Firing $R$, $S_1$ and $S_3$ we obtain the term $P=(I\,y)(I\,y)$; the
two redexes $T_3$ and $T_4$ inside $P$ looks sharable, although
they have no ancestor in common: $T_3$ is a residual of $T_1$, that in
turn was created by $S_1$, while $T_4$ has just been created by $S_3$.
In order to relate $T_3$ and $T_4$, we need to consider a {\em different}
reduction for $M$, in this case the innermost reduction of $S$
leading to $\Delta(I \, y)$ and observe that both $T_3$ and $T_4$ are
residual (w.r.t. to $R_1$) of the same redex $T$.

\begin{figure}[htb]
  \begin{center}
  \includegraphics[width=.9\columnwidth]{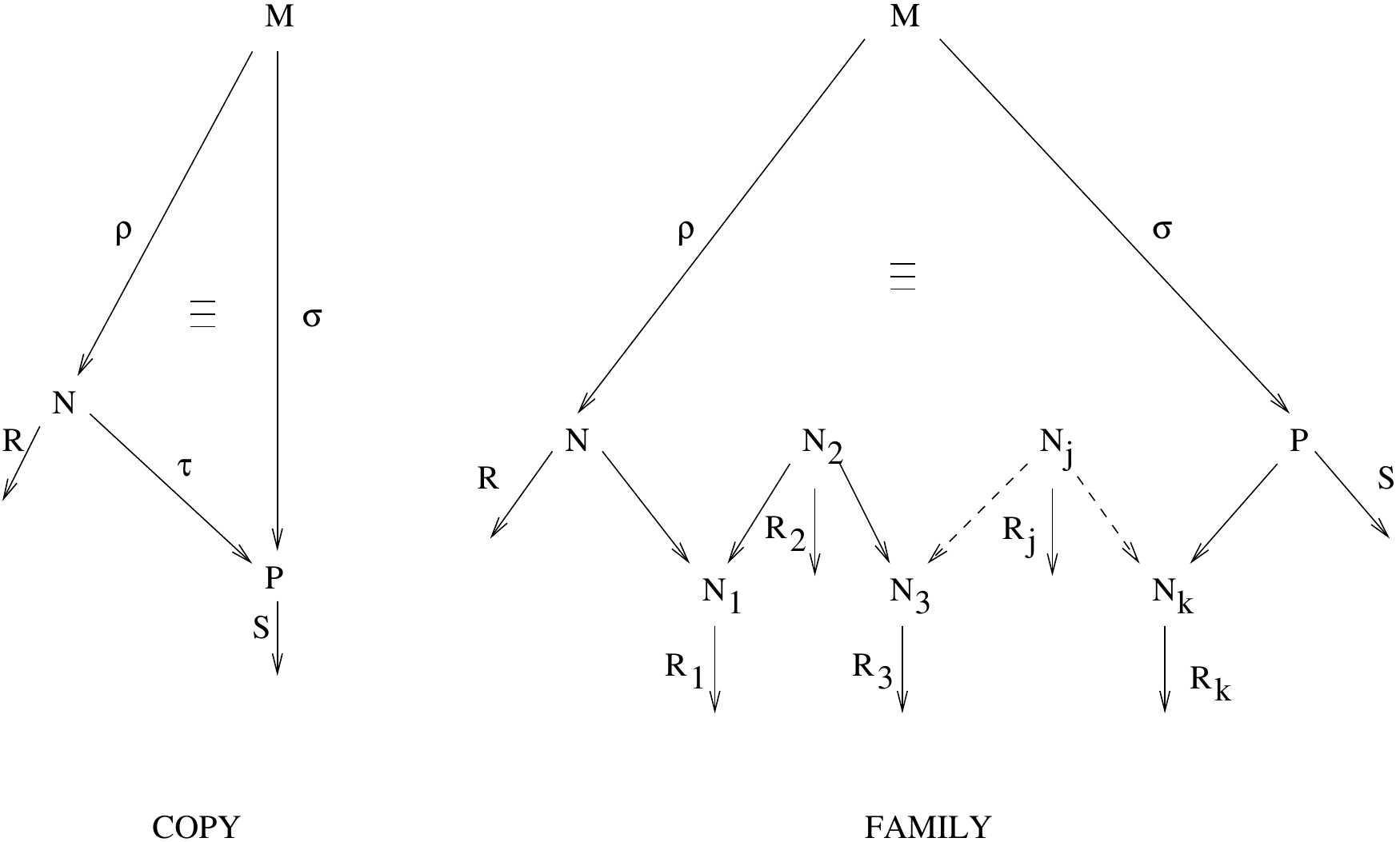}
  \caption{$\Delta = \lambda x.x\,x$, $F=\lambda z.z\,y$ and $I=\lambda x.x$}
  \label{fig:copyzigzag}
  \end{center}
\end{figure}

In general (see Figure~\ref{fig:copyzigzag}),
we say that a redex $S$ with history $\sigma$ is a {\em copy} of a redex $R$ with
  history $\rho$, written $\rho R \leq \sigma S$, if and only if there
  is a derivation $\tau$ such that $\rho \tau$ is permutation equivalent
  to $\sigma$ ($\rho \tau \equiv \sigma$) and $S$ is a residual of $R$
  with respect to $\tau$ ($S\in R/\tau$).

  The symmetric and transitive closure of the copy relation is called
  the {\em family} relation, and will be denoted with $\simeq$.

Two redexes are sharable according to L\'evy if and only if they belong
to a same family in the above sense.

It is important to observe that the family
relation is not just defined over redexes, but it is relativized
with respect to a reduction (the redex history) from some initial expression;
as a consequence we will only be able to relate redexes originated
from {\em a same} term $M$, and the choice of initial term {\em is relevant}
to determine sharing.

For instance, in the case of the example in Figure~\ref{fig:esempio}, 
if instead of start reducing from $\Delta (F\,I)$ we start from
$(F\,I)(F\,I)$ then, according to L\'evy, we loose the possibility to share 
$T_3$ and $T_4$ inside $P$.
Levy's notion aims to preserve the
sharing ``inherent'' in the initial $\lambda$-term, and not
to recognize common subexpressions generated along the
reduction (see \cite{GrabmayerR14} for an investigation of incremental 
sharing) .
Two redexes can be shared when they have been created in
essentially the same way, and not when they happen to look similar
due to ``syntactical coincidences''.

\begin{figure}[htb]
  \begin{center}
  \includegraphics[width=.8\columnwidth]{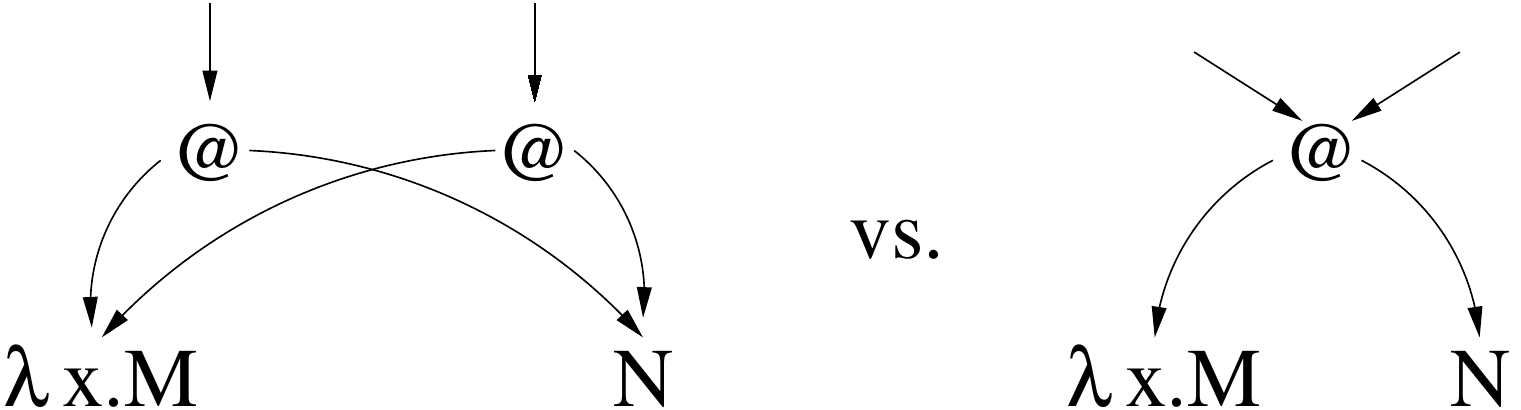}
  \caption{An example of super optimal sharing}
  \label{fig:superoptimal}
  \end{center}
\end{figure}

The critical situation is described in Figure~\ref{fig:superoptimal}.

This kind of configurations may be addressed, at some extent,
by {\em memoization} techniques: if we cash
the result of the first redex, and we meet the ``same'' 
configuration again, then we can reuse the previous result
for the second computation. The delicate point is to understand what
we mean by ``same'': intensional equality may be too restrictive, and
at the same time it may clutter the memoization table with too many terms;
on the other side, as explained in
Section~\ref{sec:hol} there is no obvious strategy to address convertibility:
in particular, the obvious approach consisting in normalizing arguments
may be in conflict with other optimality constraints (without considering the
possibility of divergence).

So, while memoization is definitely not a panacea, it is true that
in some situation can be more efficient than optimal sharing \`a la L\'evy.

A context where memoization turns out to be particularly effective is
on finite structures \cite{finite}. The advantage of working in a finite
setting is that instead of
performing memoization ``on demand'', we can work in parallel on all possible
inputs, unfolding a function into a finite vector of cases (that is,
essentially, its graph). Moreover, in this setting, types are strictly 
related to the dimension of data: this provides guidelines for the use of 
memoization, preventing to build huge hashing tables.
The resulting calculus offers an efficient framework for the
evaluation of finite terms in conjuction with a reasonably simple 
meta-theory, permitting 
a detailed and formal investigation of the complexity of reduction.

\section{Do we need higher order?}
The real question, however, is if we really need higher-order.
As a matter of fact, functional programming makes a very modest use
of it. Passing functions is used as a way to improve the parametricity
of programs, and not as a computational device. Higher order order
structures are hardly ever used as a datatype, and dynamically synthesizing
functions is much less frequent than expected. The fact that
functional languages survive without the need of optimal reduction
techniques is merely due to this fact.

The danger inherent in higher order programming is well testified
by a long series of studies relating complexity classes to hierarchies
of terms with increasing type rank
(see e.g. \cite{Gurevich83,Goerdt92prim,GoerdtS90,HillebrandK96,finite}).
For instance, even working in a restricted finite setting, terms of system $T$
of rank $2$ are already polynomially complete, and their complexity
can become rapidly unfeasible at higher ranks. 

Even the recent result in \cite{AccattoliL16} can be understood in this sense.
In order to simulate a (bounded) Turing machine you just need to
encode the transition function between configurations, that is a linear function, and
have the possibility to iterate it. On these trivial lambda terms
even a silly strategy like leftmost outermost reduction turns out to be
effective. Of course, this tells you nothing about the best
way to evaluate lambda terms. If you really want to learn a lesson
from this result is that, in order to encode Turing machines, you
do not really need the full expressive power of lambda terms, and
in particular you do not need higher-order (but to build sufficiently
large ``clocks''). This is not surprising:
in fact, to efficiently compute a Turing machine, you just need
$\dots$ a Turing machine.

\begin{acks}
  This short note was mostly motivated by a recent Haskell discussion thread
  debating {\em why isn't anyone talking about optimal lambda calculus implementations?}\footnote{
    \url{https://www.reddit.com/r/haskell/comments/2zqtfk/why_isnt_anyone_talking_about_optimal_lambda/}}. Unfortunately, the thread was already archived when I noticed
  it and did not have the opportunity to post my contribution.
\end{acks}

\bibliographystyle{ACM-Reference-Format-Journals}



\end{document}